# A note on the Nelson Cao inequality constraints in the GJR-GARCH model: Is there a leverage effect?

Stavros Stavroyiannis [(a,\*)]

[(a)] Department of Accounting and Finance, School of Management and Economics, Technological Educational Institute of Peloponnese, Greece

[(\*)] Corresponding author. Tel: (+30) 2721045303.

e-mails: computmath@gmail.com, stavroyian@teikal.gr

## Abstract

The majority of stylized facts of financial time series and several Value-at-Risk measures are modeled via univariate or multivariate GARCH processes. It is not rare that advanced GARCH models fail to converge for computational reasons, and a usual parsimonious approach is the GJR-GARCH model. There is a disagreement in the literature and the specialized econometric software, on which constraints should be used for the parameters, introducing indirectly the distinction between asymmetry and leverage. We show that the approach used by various software packages is not consistent with the Nelson-Cao inequality constraints. Implementing Monte Carlo simulations, despite of the results being empirically correct, the estimated parameters are not theoretically coherent with the Nelson-Cao constraints for ensuring positivity of conditional variances. On the other hand ruling out the leverage hypothesis, the asymmetry term in the GJR model can take negative values when typical constraints like the condition for the existence of the second and fourth moments, are imposed.

## 1. Introduction

Financial time series returns have a variety of properties that are often met, and they are named stylized facts. These include leptokurtosis, heteroskedasticity, volatility scattering, the leverage effect, and long memory. The stylized facts as well as Value-at-Risk and Expected Shortfall measures are consistently modelled within the GARCH framework of Bollerslev (1986). More advanced models include the EGARCH (Nelson, 1991), GJR (Glosten, Jagannathan, and Runkle, 1993), APARCH (Ding, Granger, and Engle, 1993), FIGARCH (Baillie, Bollerslev, and Mikkelsen, 1996a and Chung, 1999), FIEGARCH (Bollerslev and Mikkelsen, 1996), and FIAPARCH (Tse, 1998) and HYGARCH (Davidson, 2001). Distributional schemes for the likelihood estimator include the standard normal distribution (Engle, 1982), the Student-t distribution (Bollerslev, 1987), the Generalized Error Distribution (GED) introduced by Subbotin (1923) and applied by Nelson (1991), the skewed GED (Hill et al., 2008; Theodossiou, 2002; Theodossiou and Trigeorgis, 2003), the skewed t-Student distribution (Fernandez and Steel, 1998; Lambert and Laurent, 2000) hereinafter referred to as SKST, and the Pearson type IV distribution (Stavroyiannis et al., 2012; Stavroyiannis and Zarangas, 2013).

The following step in these processes is to identify the proper constraints for the parameters as well as any possible relaxation. In the initial GARCH model,

$$\sigma_t^2 = \omega + \sum_{i=1}^{p} \beta_i \sigma_{t-i}^2 + \sum_{j=1}^{q} a_j \varepsilon_{t-j}^2 \tag{1}$$

Bollerslev imposed the conditions $\omega \geq 0$, $\beta_i \geq 0$, $a_j \geq 0$, for all $i = 1 \dots p, j = 1 \dots q$ (for a detailed analysis on the constraints see Chen and An, 1998; Ling, 1999; Ling and McAleer, 2002). For simpler cases like the GARCH(1,2) model

$$\sigma_t^2 = \omega + a_1 \varepsilon_{t-1}^2 + a_2 \varepsilon_{t-2}^2 + \beta_1 \sigma_{t-1}^2 \tag{2}$$

the conditions are, $\omega \geq 0$, $0 \leq \beta < 1$, $a_1 \geq 0$, $\beta a_1 + a_2 \geq 0$, and similarly for the GARCH(2,1) model,

$$\sigma_t^2 = \omega + a_1 \varepsilon_{t-1}^2 + \beta_1 \sigma_{t-1}^2 + \beta_2 \sigma_{t-2}^2 \tag{3}$$

the conditions are relaxed via $\omega \geq 0$, $a_1 \geq 0$, $\beta_1 \geq 0$, $\beta_1 + \beta_2 < 1$, $\beta_1^2 + 4\beta_2 \geq 0$ (Nelson and Cao, 1992). Both of these cases present an admissible region where a parameter in the conditional volatility equation can assume a negative value.

Advanced GARCH models might fail to converge for a variety of reasons related to the complexity of the likelihood surface shape, especially in the case where the model and the distribution function have many parameters, or in the case of multivariate models. The algorithm used during the log-likelihood maximization might get trapped in an infinite loop over the same parameters, in the case where the surface is flat in one or several dimensions. The objective function is not well-behaved near the boundaries of the parameter's space due to the constraints set for the existence of the second and fourth moments of the distribution (for a detailed description see Silvenoinnen, 2008; Silvennoinen and Teräsvitra, 2009). Under such conditions, a usual parsimonious approach in the literature is the GJR-GARCH model.

For the GJR(1,1) case

$$\sigma_t^2 = \omega + (\alpha + \gamma I_{t-1}^-)\varepsilon_{t-1}^2 + \beta\sigma_{t-1}^2 \tag{4}$$

McAleer (*) indicates that if the residuals follow a normal distribution the condition for the existence of the second and fourth moments respectively is given by

$$\beta + a + \gamma/2 < 1 \tag{5a}$$

$$\beta^2 + 2\beta\alpha + 3a^2 + \beta\gamma + 3a\gamma + 3\gamma^2/2 < 1. \tag{5b}$$

In case of a t-Student distribution with $v$ degrees of freedom, the condition for the existence of the fourth moment is

$$\beta^2 + 2\beta\alpha + sa^2 + \beta\gamma + (2\alpha\gamma + \gamma^2)s/2 < 1 \tag{6}$$

where $s = 3(v-2)/(v-4)$, and $v \geq 5$.

When it comes to GJR modeling there is a big ambiguity and discussion in the literature about the role of the $\gamma$ parameter; it can be identified either as asymmetry or leverage. According to the Caporin and McAleer (2012) thesis, asymmetry is included to a model specification to account for the stylized fact that positive and negative shocks of equal magnitude have different impacts on volatility. On the other hand, leverage is intended to capture the possibility that negative shocks increase volatility while positive shocks decrease volatility or, as discussed in Black (1976) and Christie (1982), a negative correlation between current returns and future volatility. They conclude that the GJR-GARCH model may be asymmetric, but it is unlikely to have the leverage property since the ARCH effect must be negative, which is contrary to virtually every empirical finding in the financial econometrics literature. The next step is to decide on the constraints of the GJR process. McAleer (2014) considered a random parameter autoregressive process,

$$h_t = E(\varepsilon_t^2 | I_{t-1}) = \omega + a\varepsilon_{t-1}^2 + \gamma\varepsilon_{t-1}^2 I(\varepsilon_{t-1}) \tag{7}$$

and noticed that although at least one of $\omega, \alpha, \gamma$ must be positive, since they are the variances of three different stochastic processes all three parameters should be positive. Finally, a formal definition is given that asymmetry for GJR exists if $\gamma > 0$, while the conditions for leverage are $a + \gamma > 0$, and $a < 0$.

The purpose of this work is to look at cases where the constraints are relaxed, and indicate possible violations of the Nelson-Cao inequalities resulting to a negative conditional variance. The first case of the SP500 returns investigates the leverage issue, while the Gold returns case, which is a safe haven, investigates the asymmetry effect. We find that the leverage hypothesis violates the positivity assumption of the variance, and should be excluded from econometric software via proper modification of the constraints used. The asymmetry hypothesis does not exhibit any violations and is valid. Section 2 describes the data and the model used, and Section 3 provides the applications and offers the concluding remarks.

## 2. Econometric methodology

### 2.1. The data

As case studies we consider observations from the historical daily close values of the Standard and Poor's SP500 index from 03-Jan-1950 to 27-Dec-2016, focusing mostly on the 03-Jan-2008 to 31-Dec-2013 data range, and the Gold spot price series from 04-Jan-2000 to 27-Dec-2016. The returns are defined as the logarithmic differences of the close prices, and in cases of missing values because of holidays, or any other reason, the previous day close price is used. The stylized facts of the returns indicate the presence of skewness, leptokurtosis, and the assorted statistical tests show deviation from normality, autocorrelation and heteroskedasticity in the returns and squared returns.

### 2.2. The model

We consider the AR(1)-GJR(1,1) model where the residuals follow the SKST distribution as implemented by Lambert and Laurent (2000). The mean and the variance equations are

$$r_t = \mu + \varphi r_{t-1} + \varepsilon_t = \mu + \varphi r_{t-1} + \sigma z_t, \quad z_t \sim skst(0,1,\xi,v) \qquad (8a)$$

$$\sigma_t^2 = \omega + a\varepsilon_{t-1}^2 + \gamma \varepsilon_{t-1}^2 I(\varepsilon_{t-1} < 0) + \beta \sigma_{t-1}^2 \qquad (8b)$$

where $I$ is an indicator function taking the value 1 when $\varepsilon_t < 0$ and zero otherwise. The SKST distribution is defined as

$$f(z|\xi,v) = \frac{2}{\xi + 1/\xi} sg[\xi(\sigma z + m)|v], \qquad z < -m/s \qquad (9a)$$

$$f(z|\xi,v) = \frac{2}{\xi + 1/\xi} sg[(\sigma z + m)/\xi|v], \qquad z \geq -m/s \qquad (9b)$$

where $\xi$ is the asymmetry parameter of the distribution, defined by the ratio of the probabilities masses above and below the mode $M$, $\xi^2 = \Pr(r_t \geq M|\xi)/\Pr(r_t < M|\xi)$, $g(.|v)$ is the standardized t-Student with $v$ degrees of freedom, $m$ and $s^2$ are respectively the mean and

variance of the non-standardized SKST distribution. By definition $\xi > 0$ and in the case of $\xi = 1$ the t-Student distribution is recovered.

## 3. Results and discussion

The first issue taken under consideration is the constraints. In order to classify the $\gamma$ parameter as asymmetry, the constraint $\gamma > 0$ has to hold. This rules out the possibility that the parameter can take negative values, although the conditions $a + \gamma > 0$, $a > 0$, and $\beta > 0$ might allowed to remain. In order to classify $\gamma$ as leverage the conditions to be satisfied are $a + \gamma > 0$, and $a < 0$. The second issue is that the results of a GJR model depend highly on the software used, affected by the inclusion or exclusion of certain constraint inequalities in the programming approach of the optimization procedure. Eviews v.8.1 and OxMetrics v.7.1 allow for a negative $a$ parameter as far as $a + \gamma > 0$, while both of them including Matlab v.2014a, R language v.3.3.2 using the rugarch package (Galanos, 2015), and Gretl v.2016d allow for a negative $\gamma$ parameter as far as $a + \gamma > 0$, and $a > 0$.

### 3.1. The leverage hypothesis: the SP500 index

The leverage hypothesis holds for $a + \gamma > 0$, and $a < 0$. Although it is stated in Caporin and McAleer (2012), that this is contrary to virtually every empirical finding in the financial econometrics literature, taking into account the recent financial crises like the dotcom of the global financial crisis, it is easy to locate certain intervals in the time series that if the programmer allows, a statistically negative $a$ parameter is observed. For the SP500 index such intervals include i.e., the 04-Jan-2000 to 27-Dec-2016 range, or the 03-Jan-2008 to 31-Dec-2013 range used in this work, hereinafter referred to as SP500. Leaving aside the argument on whether or not a simple GARCH specification is appropriate in case of structural breaks, the analysis will be solely based on the software output. Table 1 presents the results of the econometric model for a variety of econometric software. The residuals follow the SKST distribution and in cases where the software didn't support one, a t-Student symmetric distribution is used. The mean is statistically significant for the Matlab and Eviews software that do not support a SKST distribution, while allowing for skewness, is not statistically significant. The autoregressive term, the GARCH, the skewness, and the tail parameters are statistically significant in all cases.

**Table 1** Results of the AR(1)-GJR(1,1) model for SP500 returns

|     | Matlab | Gretl | R | Eviews | OxMetrics |
|-----|--------|-------|---|--------|-----------|
| μ | 0.0007* | 0.0004 | 0.0004 | 0.0006* | 0.0003 |
| φ | -0.0567* | -0.0697* | -0.0698* | -0.0544* | -0.0651* |
| ω | 1.89e-06* | 2.09e-06* | 2.00E-06 | 1.72e-06* | 1.97e-06* |
| α | 1.21e-08 | 4.93e-08 | 1.00e-07 | -0.0435* | -0.0450* |
| β | 0.895* | 0.894* | 0.8940* | 0.9239* | 0.9236* |
| γ | 0.179* | 0.184* | 0.1839* | 0.2059* | 0.2133* |
| ξ | #NA | -0.1573* | 0.8547* | #NA | -0.1656* |
| v | 6.8164* | 7.493* | 7.4587* | 6.8608* | 7.438* |

*Notes: (\*) denotes statistical significance at the 5% critical level, and (#NA) indicates that a symmetric t-Student distribution is used.*

The difference in the R computing language result for the skewness is due to the fact that it calculates the value of $\xi$ which by definition is a positive number, while Gretl and OxMetrics use $\ln(\xi)$ for the practical reason to indicate the sign of the skewness. The leverage parameter $\gamma$ is positive and statistically significant in all software. The ARCH parameter is of interest since software that allows $a < 0$ result in a statistically significant negative value, where if negative values are forbidden then the parameter assumes a not statistically significant zero value. Noticing that both Eviews and OxMetrics support the negative ARCH parameter, this is evidence that is not an attribute of model misspecification using a symmetric distribution. The validity of the results has been cross-checked using native Matlab code where the constraints are tailored at will, via implementation of the Pearson type-IV as an alternative distribution.

The log-likelihood of the standardized SKST used in the optimization procedure is,

$$L_{SKST} = C - 0.5 \sum_{t=1}^{T} \left\{ \log \sigma_t^2 + (1+v) \log \left[ 1 + \frac{(sz_t + m)^2}{v-2} \xi^{-2I_t} \right] \right\} \tag{10a}$$

$$C = T \left\{ \log \Gamma \left( \frac{v+1}{2} \right) - \log \Gamma \left( \frac{v}{2} \right) - 0.5 \log[\pi(v-2)] + \log \left[ \frac{2}{\xi + \frac{1}{\xi}} \right] + \log s \right\} \tag{10b}$$

During the optimization process the variance cannot assume a negative value. The term $\log \sigma_t^2$ will approach $-\infty$ as the variance approaches zero and a negative value will never be reached. When it comes to practical applications one way to attack this problem is via Monte Carlo simulation, used by all major banks (Mehta et al., 2012) for Value-at-Risk (VaR) and related measures. We do 100 Monte Carlo AR(1)-GJR(1,1) simulations of varying time length from $T = 100$ to $T = 20000$ via the specification,

$$r_t = 0.000296 - 0.065107 r_{t-1} + \varepsilon_t = 0.000296 - 0.065107 r_{t-1} + \sigma_t u_t \tag{11a}$$

$$\sigma_t^2 = 0.019651 \times 10^{-4} - 0.045001 \varepsilon_{t-1}^2 + 0.923580 \sigma_{t-1}^2 + 0.213252 \varepsilon_{t-1}^2 I(\varepsilon < 0) \tag{11b}$$

where $u_t$ follows a standardized (zero mean and variance one) SKST with $v = 7.43802$ degrees of freedom and $\log \xi = -0.1656$ skewness parameter. The results show that in most of the Monte Carlo cases the simulation has stopped because of a negative variance, shown in Fig. 1 for a time length of 1000.

**Figure 1** Monte Carlo simulation paths exhibiting a negative conditional variance

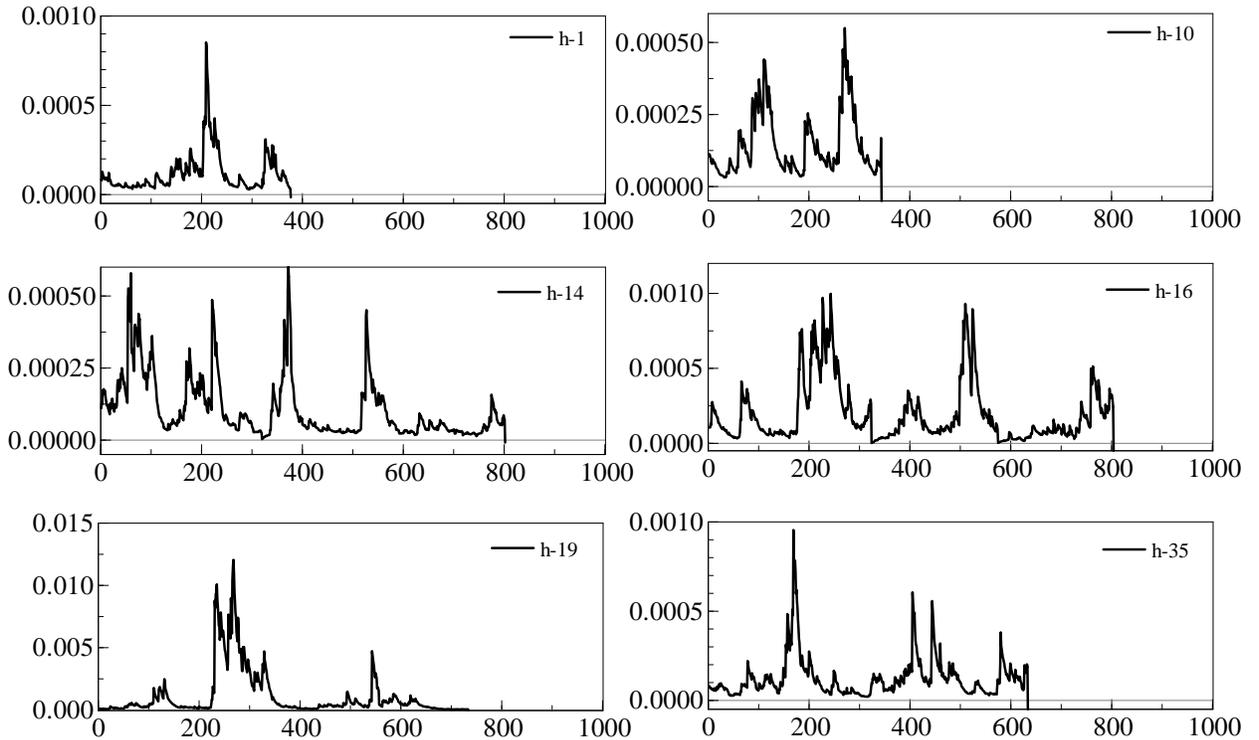

To check the success of the MC tries, we consider a survival number SR that is the number of the paths that made it to the end without encountering a negative variance. The dependence of the time length on MR (Fig. 2) shows a fast decay since given enough steps very few paths remain (the line is guide to the eye). Therefore although the leverage constraints seem to be consistent with the in-sample model specifications fitting, the approach is not correct and the constraint $a > 0$ should always be considered.

**Figure 2** Dependence of SR as a function of simulations length T

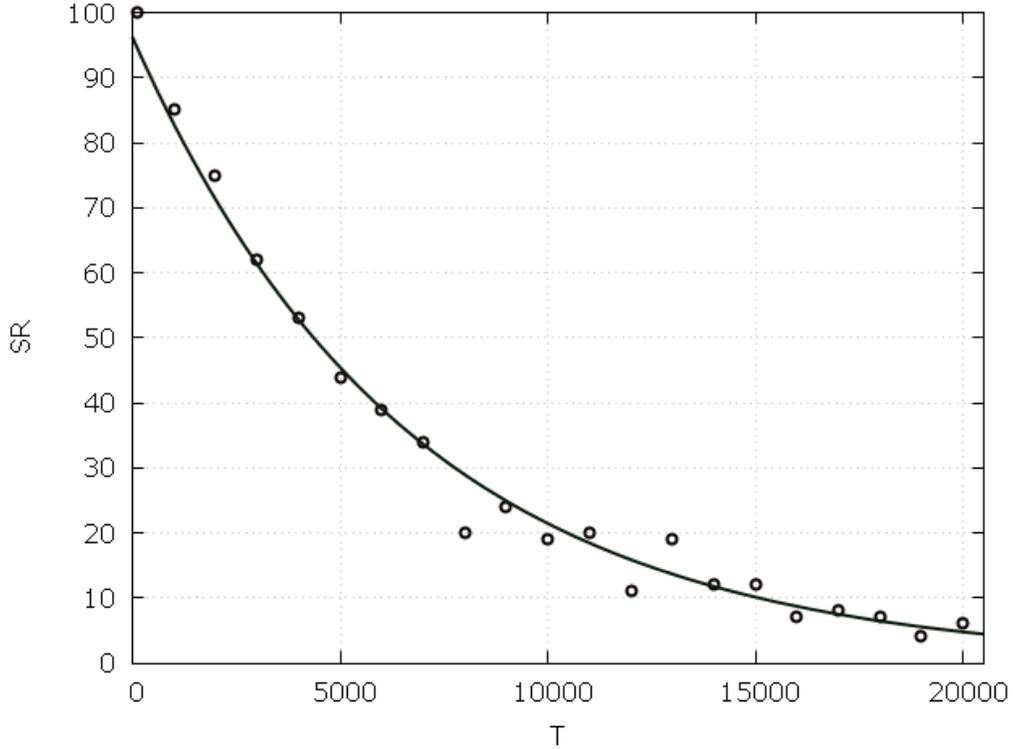

Working around the initial model with an ARMA(1,1)-GJR(1,1) specification,

$$r_t = 0.000378 + 0.460538 r_{t-1} - 0.533473 \varepsilon_{t-1} + \varepsilon_t \tag{12a}$$

$$\sigma_t^2 = 0.018770 \times 10^{-4} - 0.039250 \varepsilon_{t-1}^2 + 0.924087 \sigma_{t-1}^2 + 0.197914 \varepsilon_{t-1}^2 I(\varepsilon < 0) \tag{12b}$$

$$\log \xi = -0.174414 \tag{12c}$$

$$v = 7.545299 \tag{12d}$$

an ARMA(1,1)-GJR(2,1) specification,

$$r_t = 0.000379 + 0.458079r_{t-1} - 0.531676\varepsilon_{t-1} + \varepsilon_t \tag{13a}$$

$$\sigma_t^2 = 0.018099 \times 10^{-4} - 0.036587\varepsilon_{t-1}^2 + 0.971335\sigma_{t-1}^2 - 0.044824\sigma_{t-2}^2$$

$$+0.188975\varepsilon_{t-1}^2 I(\varepsilon < 0) \tag{13b}$$

$$\log \xi = -0.174029 \tag{13c}$$

$$v = 7.547291 \tag{13d}$$

and an ARMA(1,1)-GJR(1,2) specification

$$r_t = 0.000409 + 0.508983r_{t-1} - 0.588970\varepsilon_{t-1} + \varepsilon_t \tag{14a}$$

$$\sigma_t^2 = 0.024810 \times 10^{-4} - 0.079038\varepsilon_{t-1}^2 + 0.058260\ \varepsilon_{t-2}^2 + 0.898783\sigma_{t-1}^2$$

$$+0.124897\varepsilon_{t-1}^2 I(\varepsilon < 0) + 0.074442\varepsilon_{t-2}^2 I(\varepsilon < 0) \tag{14b}$$

$$\log \xi = -0.175171 \tag{14c}$$

$$v = 8.253047 \tag{14d}$$

the conclusion about the leverage hypothesis is not refuted; the negative ARCH effect remains statistically significant and all specifications fail to produce reliable simulations without encountering a negative variance. The first part of our work confirms that no leverage effect is possible for the GJR(1,1) case.

## 3.2. The asymmetry hypothesis: the Gold spot price

Baur & Lucey (2010), and Baur and McDermott (2010), identified three asset classes and provided a formal distinction and definition of hedge, diversifier, and safe haven. A safe haven is an asset that is uncorrelated on the average with a portfolio but during crises it exhibits a negative correlation. Gold is a commodity exhibiting interesting features towards the behavior of the investors during crises since it possesses the properties of a safe haven. The results of the AR(1)-GJR(1,1) specification is given in Table 2. Since $a + \gamma > 0$, $a > 0$ all software allow for a negative value of the $\gamma$ parameter. Running native code with $a + \gamma > 0$, $a > 0$, $\gamma > 0$, the parameter is practically zero ($\gamma = 5.1193e - 09$) with t-statistics 0.011.

**Table 2** Results of the AR(1)-GJR(1,1) model for the Gold spot returns

| GJR(1,1) | Matlab | Gretl | R | Eviews | OxMetrics |
|---|---|---|---|---|---|
| $\mu$ | 0.0004* | 0.0004 | 0.0004* | 0.0004* | 0.0004* |
| $\varphi$ | -0.04* | -0.04* | -0.04* | -0.04* | -0.04* |
| $\omega$ | 1.46e-06* | 1.4e-06* | 1.0e-06 | 1.4e-06* | 1.4e-06* |
| $\alpha$ | 0.08 | 0.08* | 0.08* | 0.08* | 0.08* |
| $\beta$ | 0.9333* | 0.9333* | 0.9334* | 0.9334* | 0.9332* |
| $\gamma$ | -0.0047* | -0.047* | -0.047* | -0.047* | -0.047* |
| $\xi$ | #NA | 0.0057 | 1.00* | #NA | -0.005 |
| $v$ | 5.22* | 5.22* | 5.23* | 5.23* | 5.219* |

*Notes: (\*) denotes statistical significance at the 5% critical level, and (#NA) indicates that a symmetric t-Student distribution is used.*

The international indices are increasing throughout the decades, as a result of the technology and "the Internet of Things", but in cases of crises there are significant losses. This effect creates the usual positive $\gamma$ asymmetric term. In the case of a crisis a safe haven asset reverses this effect, since there is a movement from the asset or portfolio towards the haven, resulting in a negative $\gamma$ parameter. In support of this reasoning we look at the dynamic properties of the SP500-Gold pair from 04-Jan-2000 to 27-Dec-2016 which includes several crises. Using a Vector Autoregressive (VAR) model on both series returns, a short-run Granger causality is identified from SP500 towards Gold and not vice-versa, as shown in Table 3.

**Table 3** VAR Granger Causality/Block Exogeneity Wald Tests

| Dependent variable: DLGOLD | | | |
|---|---|---|---|
| Excluded | Chi-sq | df | Prob. |
| DLSP | 21.762 | 2 | 0.0000* |
| All | 21.762 | 2 | 0.0000 |

| Dependent variable: DLSP | | | |
|---|---|---|---|
| Excluded | Chi-sq | df | Prob. |
| DLGOLD | 4.227 | 2 | 0.1208 |
| All | 4.227 | 2 | 0.1208 |

*Notes: (\*) denotes statistical significance at the 5% critical level*

This shows that a negative asymmetry parameter has a solid financial ground. Assuming that $a > 0$ and $\beta > 0$ by choosing an appropriate grid with positive values, the admissible region of the inequalities $a + \gamma > 0$ and $\beta + a + \gamma/2 < 1$ considering a normal distribution (Eq. 5a) is shown in Fig. 3, allowing the asymmetry parameter to take negative values.

**Figure 3** Admissible region of the inequalities $a, \beta > 0$, $a + \gamma > 0$, and $\beta + a + \gamma/2 < 1$

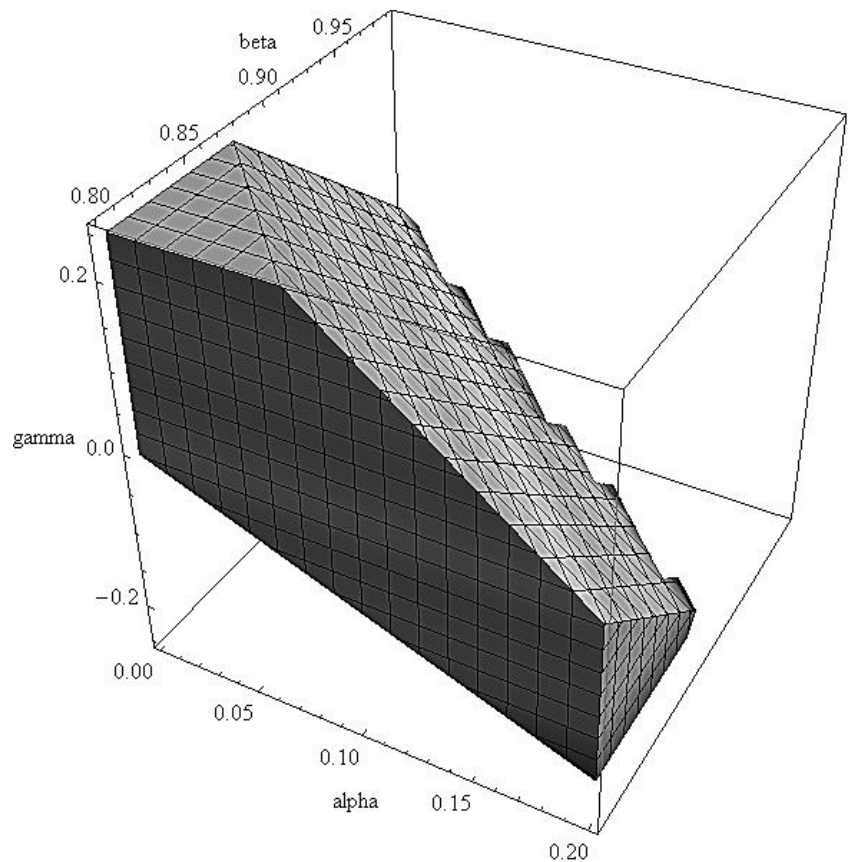

In case of a t-Student distribution with the constraint of Eq. (5b) the region does not change much and the possibility that the asymmetry parameter can take negative values remains. Using a barrage of Monte Carlo simulations with varying path lengths, the conditional variance remains positive; therefore, the asymmetry hypothesis for the GJR(1,1), as far as the constraints $a > 0$ and $a + \gamma > 0$ are valid, cannot be refuted.

## 4. Conclusion

This paper examined the strength of the constraints used in several software for the optimization of the GJR(1,1) model. Relaxing certain constraints in the GJR(1,1) case can characterize the $\gamma$ parameter as of asymmetry or leverage. The results showed that leverage cannot exist since it leads to violation of the Nelson-Cao inequality constraints and the positivity property of variance. On the other hand, under certain assumptions on the constraints the asymmetry hypothesis does not lead to any inconsistent results.